\RequirePackage{fix-cm}
\documentclass[smallextended]{svjour3}       % onecolumn (second format)
\smartqed  % flush right qed marks, e.g. at end of proof
\usepackage{graphicx}
\begin{document}

\title{$\rm\Lambda$CDM
}
\subtitle{Much more than we expected, but now less than what we want}


\author{Michael S. TURNER        
%% \and
 %%       Second Author %etc.
}


\institute{M.S. Turner \at
             Kavli Institute for Cosmological Physics \\
             The University of Chicago\\
             5640 S. Ellis Avenue\\
             Chicago, IL  60637-1433\\
             USA \ \ \ \
              %%Tel.: +123-45-678910\\
             %% Fax: +123-45-678910\\
              \email{mturner@uchicago.edu}           %  \\
          %% \and
           %%S. Author \at
              %%second address
}

%%\date{Received: date / Accepted: date}
\date{Proceedings of the Lemaitre Workshop on Black Holes, Gravitational Waves and Space Time Singularities, published in Foundations of Physics (2018)}

\maketitle

\begin{abstract}
The  $\rm\Lambda$CDM cosmological model is remarkable:  with just 6 parameters it  describes the evolution of the Universe from a very early time when all structures were quantum fluctuations on subatomic scales to the present, and it is consistent with a wealth of high-precision data, both laboratory measurements and astronomical observations.  However, the foundation of $\rm\Lambda$CDM involves physics beyond the standard model of particle physics:  particle dark matter, dark energy and cosmic inflation.  Until this `new physics' is clarified, $\rm\Lambda$CDM is at best incomplete and at worst a phenomenological construct that accommodates the data.  I discuss the path forward, which involves both discovery and disruption, some grand challenges and finally the limits of scientific cosmology.
%%Include keywords, PACS and mathematical subject classification numbers as needed.
%%\keywords{First keyword \and Second keyword \and More}
\end{abstract}

\section{The ${\rm\Lambda}$CDM paradigm}
\label{sec:1}
\subsection{Some history} For me, cosmology began in the late 1970s.  Then,  the Cosmic Microwave Background (CMB) was well established but only the dipole anisotropy had been measured.  The redshifts of a few thousand galaxies had been determined, and a high redshift galaxy was $z \sim 0.3$ and the QSO record holder was $z \sim 3.7$.  CCD cameras were just entering the astronomical scene,  $H_0$ was either 50 or 100 km/s/Mpc, each with tiny claimed error bar, and  the best bet value for $\Omega_0$ was $\sim 0.1$. On the theoretical side, big-bang nucleosynthesis (BBN) was  well established and stood as the earliest cosmological outpost at about $10^{-2}\,$sec.  Gravitational instability was the default model for structure formation, but there was no evidence for the seed density perturbations or any compelling ideas about their origin.  There were two  schools of thought on structure formation:  Zel'dovich's adiabatic (now called curvature) school that favored the scale-invariant Harrison spectrum of primeval fluctuations with structure forming `top-down' and Peebles' isothermal (now called isocurvature) school with structure forming `bottom-up.' 

While cosmology is a young science, it dates back further than the 1970s!  On the theoretical side, the starting point is Einstein's General Relativity (1915):  without it, it is not possible to sensibly discuss the Universe and interpret the observations. Hubble's two big contributions followed:  establishing the existence of other galaxies in 1925 and establishing the relationship between redshifts and distances  in 1929.  While the current interpretation of the Hubble diagram as the expansion of space and redshifting of light seems so straightforward, it took a good ten years to arrive at such a consensus and involved much confusion.\footnote{I think it is fair to say that it took almost 100 years to fully understand General Relativity.  Years ago, Chandrasekhar captured the enormity of the task in a quip to an experimental colleague, ``you see Bruce, Einstein's understanding of General Relativity was very limited,'' which evoked a loud laugh.}  The work of Lemaitre, deSitter and others was crucial to establishing our current understanding of the expansion, and Kragh has discussed Lemaitre's seminal contributions at this meeting \cite{Kraghe}.

Shortly before his death in 1953, Hubble summarized the state of observational cosmology \cite{Hubble1953}.  Between the Lick program led by Mayall and the Mt. Wilson/Palomar program led by Humason, just over 800 galaxy redshifts had been measured \cite{Humason1956}.  For field galaxies,  redshifts were $z \le 0.04$, with a few cluster galaxies at $z \sim 0.2$.  The best estimate for the Hubble constant was 180 km/s/Mpc.  Shortly thereafter, in two influential papers \cite{SandagePhysToday,Sandage200inch}, Allan Sandage characterized cosmology as the quest for two numbers\footnote{Most younger cosmologists today would be hard pressed to define $q_0$, and in fact, it is not measurable \cite{Neben}!} -- $H_0$ and $q_0$ -- and argued that they would be determined with the 200-inch Hale telescope at Mt. Palomar.  The discovery of the CMB in 1965 and its interpretation as evidence for a hot beginning changed everything:  Before galaxies, there was a radiation-dominated phase where the light elements were synthesized (from $10^{-2}\,$sec to $200\,$sec) and the oldest light originated (last scattering of the CMB at $380,000\,$yrs and $z \simeq 1100$).

This brings us to the late 1970s, but not  to $\rm\Lambda$CDM.  The big change in the 1980s was the realization of the profound connections between quarks and the cosmos.  The discovery that the fundamental constituents are point-like quarks and leptons with perturbatively-weak interactions opened up the very early Universe, times earlier than $10^{-6}\,$sec, for exploration \cite{TheEarlyUniverse}.  Speculations about the earliest moments, motivated by ideas about unification of the particles and forces,  changed cosmology forever and led to the ideas underpinning $\rm\Lambda$CDM and modern cosmology. 

The first shift was to view the large photon (or entropy) to baryon ratio (around 2 billion) as a small net baryon number to photon ratio that arose dynamically due to $B$, $C$ and $CP$ violating processes that occurred out of thermal equilibrium (baryogenesis).  Next, the dark matter problem of cosmology became an opportunity for particle physics:  stable particles (beyond baryons) left over from the quark soup beginning  account for the bulk of the mass density of the Universe.  

Inflation, a period of accelerated expansion caused by a slowly-evolving scalar field initially displaced from the minimal of its scalar potential, was proposed in 1981 to explain the smoothness of the Universe, the absence of magnetic monopoles and other unwanted relics as well as explaining the small density perturbations needed to seed the formation of structure in the Universe \cite{Guth}.  

The final piece of current paradigm, dark energy, was added most recently.  It was proposed in 1984, as a smooth, highly-elastic form of energy that would not interfere with the formation of structure and would account for 70\% of the critical density.  It was needed because because dark matter was coming up short of the critical density necessary for the flat Universe predicted by inflation \cite{TSK}.  By virtue of its highly elastic nature -- equation-of-state $w\simeq -1$ -- dark energy has repulsive gravity and leads to accelerated expansion.  That accelerated expansion was discovered in 1998, as discussed by Filippenko \cite{Filippenko} and Suntzeff \cite{Suntzeff}.

The deep connections between the quarks and the cosmos go far beyond the fact that the early Universe was comprised of quark soup, and they have led to a paradigm shift in cosmology and the coming together of the study of the very big (cosmology) and the very small (particle physics).  That profound change of view is embodied both in $\rm\Lambda$CDM and the big questions we are asking about the Universe today that go far beyond the simple kinematics of $H_0$ and $q_0$ (see \S\ref{sec:4}).  

The three major features of  $\rm\Lambda$CDM have testable consequences that have driven observational cosmology since the 1980s:  1.  Inflation  predicts a flat (critical density) Universe, with almost scale-invariant, Gaussian density perturbations and gravitational waves; 2. Cold dark matter (slowly-moving, dark-matter particles), together with the inflationary density perturbations, leads to a very detailed picture of structure formation -- the cold dark matter paradigm (CDM); and 3. $\rm\Lambda$ as the dark energy has a host of consequences and can be readily falsified.  

\begin{figure}
\includegraphics[width=0.9\textwidth]{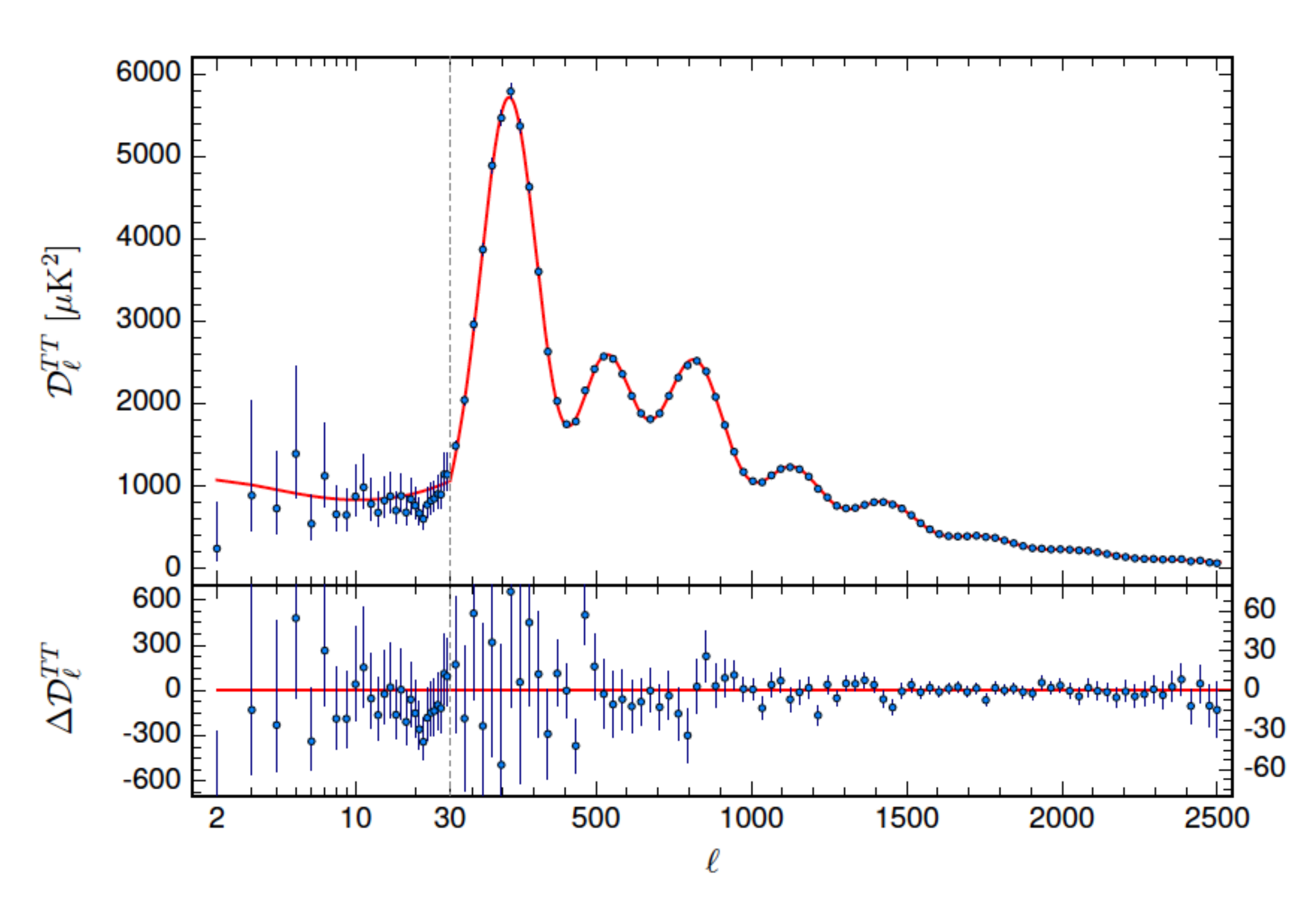}
\includegraphics{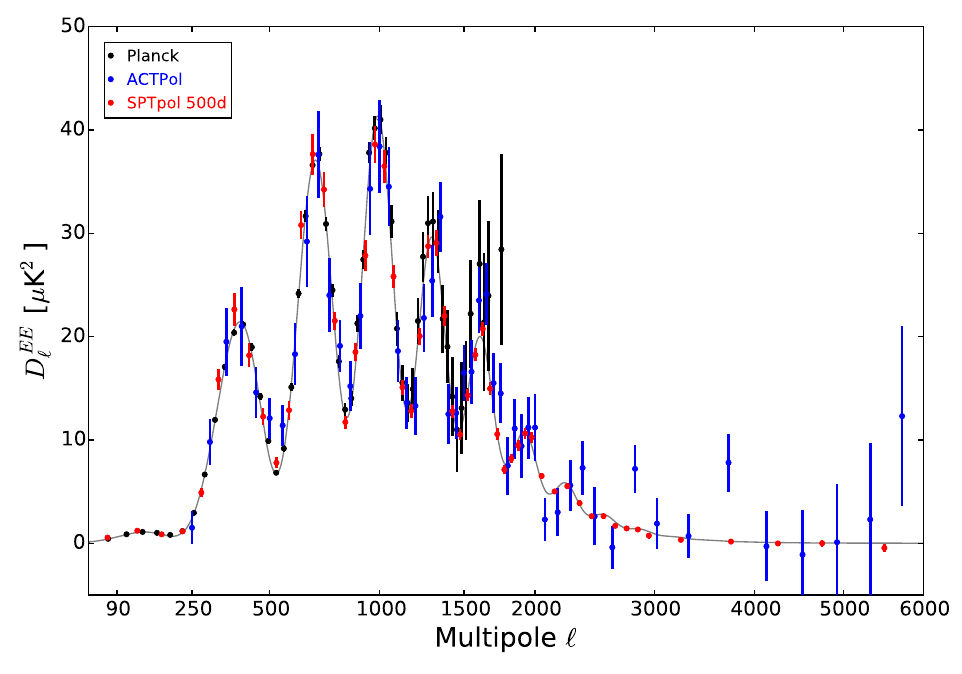}
\caption{CMB anisotropy as measured by Planck (upper) and polarization as measured by Planck, SPT and ACTPol (lower); in both panels, the curve is the best-fit $\rm\Lambda$CDM model.}
\label{fig:1}       % Give a unique label
\end{figure}

After three decades of intense testing $\rm\Lambda$CDM is consistent with an enormous body of diverse, high-quality observational data, from measurements of large-scale structure to the anisotropy of the CMB, from light-element abundances to supernovae determinations of the expansion rate, as well as a host of laboratory measurements.  $\rm\Lambda$CDM is characterized by just six numbers \cite{Planck2013}:
\begin{enumerate}
\item {Baryon density:} $\omega_B \equiv \Omega_B h^2$
\item  {Matter density:}  $\omega_M \equiv \Omega_M h^2$
\item {Spectral index density perturbations:} $n_S$ 
\item Amplitude of density perturbations:  $A_S$ 
\item {Optical depth to the CMB due to re-ionization:}  $\tau$ 
\item  {Sound horizon at recombination:}  $\theta_{MC}$
\end{enumerate}

The excellent fit to the CMB anisotropy data is shown in Fig.~1.  Fitted values for these parameters, which include $\omega_B = 0.0222 \pm 0.00023$, $\omega_M = 0.142\pm 0.0022$, $\tau = 0.078\pm 0.019$ and $n_S = 0.965 \pm 0.006$,  indicate the percent-level precision that is now being achieved.  From these (and other cosmological data) one can derive the following set of cosmological parameters that describe our Universe:
\begin{eqnarray}
T_0 & = & 2.7255 \pm 0.0006\,{\rm K} \\
t_0 & = & 13.80 \pm 0.02\,{\rm Gyr}\\
\Omega_0 & = & 1.00 \pm 0.005 \\
\Omega_M & = & 0.316 \pm 0.01 \\
\Omega_B & = & 0.048 \pm 0.001 \\
\Omega_\Lambda & = & 0.685 \pm 0.01 \\
H_0 & = & 67.3 \pm 0.7\,{\rm km/s/Mpc}
\end{eqnarray}

Beyond its impressive precision $\rm\Lambda$CDM describes in detail the Universe from quantum fluctuations to quark soup to neutrons and protons and then nuclei, from the recombination of ions and electrons and photon last scattering to the growth of structure and the formation of astrophysical objects that can be studied today with telescopes -- stars, galaxies, clusters of galaxies, voids, walls and superclusters.  

In particular, $\rm\Lambda$CDM provides the background space-time and initial conditions for what I call `astrophysical cosmology,' namely, the story of the Universe from lumpy, ionized gas to the formation of the  stars, galaxies and large-scale structure.  With the remarkable computing power that is being brought to bear and the powerful array of instrumentation and telescopes that allow us to probe the Universe back to redshift $z\sim 10$ and higher,  progress in reconstructing the history of the Universe from the first stars until today is equally impressive.  For me, this success is best summarized by the Madau plot of the star formation history of the Universe; see Fig.~2.

In 1978 I would never have imagined that we would find ourselves having progressed so far in our knowledge about and understanding of the Universe.  While I cannot speak for Lemaitre, Hubble, Einstein and Sandage, I suspect that they would be similarly surprised and awed.  $\rm\Lambda$CDM is a remarkable achievement, one that has raised our expectations even higher.

\begin{figure}
\includegraphics[width=0.7\textwidth]{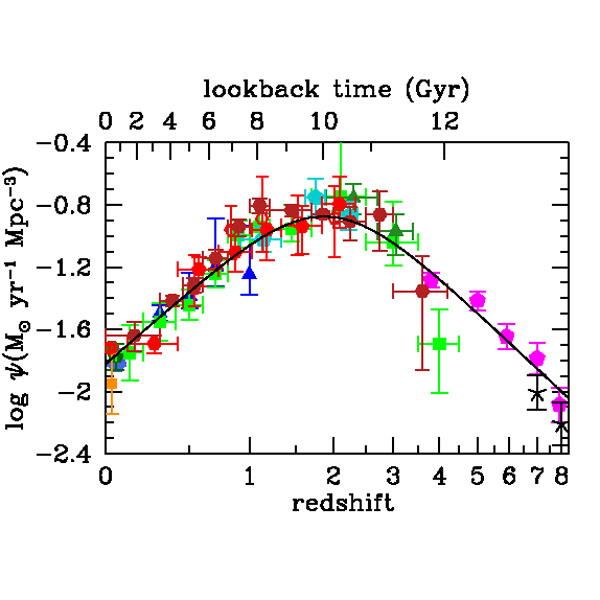}
\caption{Madau plot of the star formation rate \cite{MadauDickinson}.}
\label{fig:2}       % Give a unique label
\end{figure}

\section{The missing physics: discovery opportunities or surprises ahead?}
\label{sec:2}

So why isn't a model that agrees with a wealth of observational data, takes us back to within a tiny, tiny fraction of a second of the beginning, and explains the Universe that we see today not enough?  Just three words:  dark-matter, dark-energy and inflation.  They are the three pillars of $\rm\Lambda$CDM, and each involves physics beyond what we know from laboratory experiments --  `physics beyond the standard model of particle physics.'  Science is much more than a list of facts or curve-fitting formulae; it is at its best a deeper understanding of the physical world.  The success of $\rm\Lambda$CDM has left us wanting more, asking  deeper questions, and expecting answers!  In science no good deed goes unpunished:  answering questions usually leads to a new set of more profound questions.

Of course, there is another way to look at $\rm\Lambda$CDM's success:  dark matter, dark energy and inflation are simply proxies for the real, even more revolutionary physics needed to describe our Universe; some have even called them epicycles.  For example, dark energy and dark matter could use be patches for the shortcomings of General Relativity.  If this is correct, the path ahead in cosmology will be even more exciting.

The conservative view today is that dark matter, dark energy, and inflation are simply physics beyond the standard model waiting to be clarified and understood:  a new particle, a new form of elastic energy and a new scalar field.  Other than neutrino mass, they are the only phenomena that cannot be explained by the standard model.\footnote{Baryogenesis is also likely in this category; I have not included it though, because there is $B$ and $CP$ violation within the standard model and some still hold out the hope that the baryon asymmetry of the Universe can be explained by the standard model or a minimal extension thereof.}

\subsection{Dark Matter}  The evidence for non-baryonic dark matter began to develop in the 1980s with the emerging gap between the baryon density determined by BBN ($\Omega_B \leq 0.1$) and the total matter density determined by a variety of astrophysical methods ($\Omega_M \geq 0.2$).  Today, there is an airtight case based upon the 50$\sigma$ discrepancy between the BBN/CMB determination of the baryon density, $\omega_B = 0.0222 \pm 0.00023$, and the CMB/LSS determination of the matter density, $\omega_M = 0.142\pm 0.0022$.  We have additional astrophysical information about the dark matter as well:  it interacts weakly  with ordinary matter and is slowly moving (`cold').  And we know that a small amount of the dark matter -- somewhere between 0.1\% and 1\% of critical density -- exists in the form of fast-moving (`hot') relic neutrinos.

On the theoretical side, attempts to go beyond the standard model of particle physics to unify the forces and particles of Nature have led to two compelling ideas for the dark matter particle:  the neutralino, the lightest superpartner in supersymmetric extensions of the standard model (mass 100\,GeV to several TeV), and the axion, a very light particle (mass of order $10^{-6}\,$eV), which is a consequence of the most attractive idea about solving the strong $CP$ problem of QCD.  There is a more generic version of the neutralino -- the WIMP for Weakly-Interacting Massive Particle -- whose motivations trace to the fact that a stable particle whose annihilation cross section is `weak,' more specifically $<\sigma v>_{\rm ann} \sim 3 \times 10^{-26}\,{\rm cm^3 sec^{-1}}$, will have the relic abundance needed to be the dark matter (sometimes called the WIMP miracle).

On the experimental side, there is a vigorous, three-pronged approach to identifying the dark matter particle:  direct detection of dark matter particles in our halo, indirect detection (detection of annihilation products from halo dark-matter annihilations) and accelerator production.  Direct-detection experiments are now probing the theoretically favored parts of the parameter space for neutralinos and axions; for indirect detection and accelerator production the same is true for WIMPs and neutralinos.

This is moment of truth for the particle dark matter hypothesis:  we have an airtight case for this bold hypothesis, a broad and sensitive-enough experimental effort to test the hypothesis, and no compelling evidence for particle dark matter -- yet.   

There are several possible futures, all interesting, though some less satisfying.  First, a detection could be right around the corner!  Second, our compelling ideas could be just that -- compelling, but not correct!  Third, there are new ideas, including asymmetric dark matter, low-mass dark matter (1\,GeV and down to tiny fractions of an eV), and the most expansive idea yet, a whole dark sector, one with its own particles and interactions, to be discovered.  It could be that rather than being close to a solution, we have just seen the tip of the iceberg of a whole new sector that interacts only very weakly with the standard model particles that we are familiar with.

It should also be said that none of the astrophysical and cosmological evidence for dark matter requires it to have other than gravitational interactions with standard model particles.  Thus, it could be that detection of the dark matter particles is far beyond our current capabilities and for the foreseeable future particle dark matter will just be an economical description of all the dark-matter phenomena.  

Finally, it is possible that the dark matter problem is actually telling us that Newton's theory -- recall, most of the astrophysical evidence for dark matter only involves Newtonian physics -- is not correct on scales much larger than the solar system, where it is well tested.  (Of course, General Relativity would have to be modified as well.) There have been and are a host of ideas, starting with Milgrom's MOND in 1983 \cite{Milgrom} and extending to string-inspired ideas (e.g., those of Verlinde \cite{Verlinde}) about how this might be.  I must confess, I find none of these ideas as compelling as the simple hypothesis of particle dark matter.  Of course, Nature may not have chosen what I find what I find compelling!

%%While I am hoping that the solution to the dark-matter problem will reveal itself soon, all four futures  are exciting.  
%%And remember, if history is any guide, just when we think we have Nature cornered, she can throw us a curve ball and send us back to the drawing board.

\subsection{Dark Energy}  The history of dark energy and cosmic acceleration is more recent, but its roots in the form of $\rm\Lambda$ go back to deSitter, Einstein, Lemaitre and Eddington and the beginning of cosmology.  $\rm\Lambda$ was invoked again by Bondi, Hoyle and Gold in their steady state cosmology in 1948 and in the 1960s to explain the great abundance of QSOs at redshift $z\sim 2$.  In the 1980s cosmic acceleration returned in the form of inflation, and beginning in 1984, $\rm\Lambda$ returned as a means to achieving a flat Universe with a matter density of order 30\% of the critical density \cite{TSK}.  

One can't discuss $\rm \Lambda$ without mentioning the greatest embarrassment of theoretical physics:  the mis-prediction of quantum vacuum energy, which is mathematically equivalent to $\rm\Lambda$.  The quantum zero-point energies, when summed over all modes, diverges; this is a fact that Pauli first realized in the 1930s.  Truncating the sum at 1\,TeV, leads to a value that is 55 orders-of-magnitude larger than what is required to explain cosmic acceleration.  For many years this big problem -- known as the cosmological constant problem -- was ignored, with the tacit assumption that the correct answer would eventually turn out to be zero and theorists just needed to figure out how.  The discovery of accelerated expansion in 1998 brought the cosmological constant problem front and center and made it a much richer and more urgent question:  a small vacuum energy could be the explanation; or something else.  One problem or two (or more)!

%%Because of its checkered history and the big, big puzzle of why $\rm \Lambda$ is so tiny many -- including me -- were surprised in 1998 when the SNeIa data from the two teams showed that the Universe really is accelerating.

The term dark energy refers to a smooth component of energy density with an equation-of-state $w$ sufficiently negative to cause the expansion to accelerate ($w< -1/3$); it was introduced to remind astronomers (and others) that cosmic acceleration might well involve more than just $\rm\Lambda$ \cite{Stromlo}.   Because of its familiarity and its mathematical equivalence to vacuum energy, $\rm\Lambda$ is the `simplest' example of dark energy, with $w\equiv -1$.  Moreover, the data are consistent with this:  $w = -1.00^{+0.04}_{-0.05}$ (sys+stat) \cite{DES1}.  There is even theoretical motivation:  some argue that the landscape of string theory, with its more than $10^{500}$ vacua, all with different values of the vacuum energy, can accommodate the seemingly small value we find \cite{Linde}.  The next level of complexity, involves the idea of a time-varying vacuum energy, implemented by a slowly-rolling, ultra-light scalar field.  While General Relativity + dark energy can accommodate the observed accelerated expansion, it could be that there is no dark energy and that the correct gravity theory will explain cosmic acceleration. 

Unlike dark matter, where there are compelling theoretical ideas and a robust experimental program to test them, the situation here is  different:  No compelling theoretical idea, and $\rm\Lambda$, with and all its baggage, fits the data well (better than 5\%).  For the moment, $\rm\Lambda$ serves as a  good proxy for the real explanation.  Moreover, with the observational program in place, there are opportunities for surprises and new physics ahead.

\subsection{Inflation}  Inflation is the most important idea in cosmology since the big bang itself.  It offers a compelling explanation for the origin of the smoothness of the Universe as well as the small inhomogeneities that seed cosmic structure, and  with particle dark matter, it has given us the highly successful CDM paradigm for structure formation.  There is evidence for two of its three basic predictions:  flat Universe and almost scale-invariant Gaussian density perturbations.  Further, heroic efforts are being mounted to test its third big prediction, an almost scale-invariant spectrum of gravitational waves \cite{Kallosh}.
%%, by searching for the $\leq 100$\,nK B-mode polarization signature they imprint of the CMB sky \cite{Kallosh}.  
Lastly, the discovery of the Higgs has provided some motivation for the scalar field(s) that are used in most models of inflation.

All that being said, inflation is not yet a well-formulated, complete theory, and there is no standard model of inflation.  At best, we have a  rudimentary description of inflation.  An analogy I like, is the one to superconductivity:  Ginzburg-Landau theory vs BCS theory.  At this stage we have a Ginzburg-Landau description, one with three robust predictions.  

There are also important unresolved issues:  the initial conditions for inflation; if it predicts the multiverse, how is one to interpret the multiverse and the testing of inflation?; and with regard to the smoothness/horizon problem, it is at best a postponement since inhomogeneity is merely pushed outside the horizon, eventually coming back inside the horizon (unless the expansion of the Universe continues to accelerate).  Some believe that these issues are telling us that inflation is not bold enough \cite{Penrose}.

While the simplest models of inflation (single-field) do an excellent  job of fitting the extant data -- flat Universe with Gaussian, almost scale-invariant density perturbations -- at this point, the best we can say is that inflation is a compact description of the earliest moments that dynamically sets some of the most fundamental features of the Universe.  The opportunities for a deeper understanding of the beginning of the Universe as well as big surprises are great.

\section{Moving forward: discovery and disruption}
\label{sec:3}
I see cosmology moving forward in two distinct ways:  discovery and disruption.  Both are made possible by the growing number of precision measurements and correspondingly precise theoretical predictions. Some of the datasets being assembled are large and complex, and often the cross correlations between large datasets will be crucial in moving forward.

Disruption and discovery are not mutually exclusive.  While disruption refers to upsetting the  $\rm\Lambda$CDM paradigm, a disruptive measurement could  also lead to a discovery.  A discovery could lead to clarifying and establishing the current paradigm, or disrupting it.  I will illustrate with a few examples of discovery and disruption that I see on the horizon.

\subsection{Inflationary B-modes:  discovery}  The third key prediction of inflation is a spectrum of almost scale-invariant gravitational waves.  The amplitude of the spectrum depends upon the energy scale of inflation, and thus the detection of these gravitational waves would reveal the epoch of inflation.  The most promising means of detection is their imprint on the CMB, in the form of the odd-parity mode (B-mode) of CMB polarization, which is {\em not} excited by scalar (density) perturbations.  

But it won't be easy; current limits to the B-mode polarization from inflation, tensor-to-scalar ratio $r < 0.1$, correspond to a B-mode signal that is less than $100\,$nK in amplitude.  B-mode polarization can be produced from E-mode (even parity) polarization arising from density perturbations by the gravitational lensing effects of large-scale structure along the line of sight.
%%Gravitational lensing of even-parity polarization (E-mode), arising from density perturbations, by intervening large-scale structure produces B-mode polarization.  
Further, foreground dust emission is polarized with roughly equal E-mode and B-mode components.  Finally,  there is no compelling theoretical prediction for the inflationary B-mode amplitude -- worst yet, some theorists say it will be undetectably small. 

In spite of all this, there is an impressive and growing effort to reach a sensitivity of $r \leq 10^{-3}$ \cite{Kallosh}.  That effort involves multi-wavelength measurements to separate foreground emission from CMB contributions; measurements on angular scales from more than one degree to a tenth of a degree (multipole ${\ell}$ from 10's to a few 1000); and the use of powerful analysis techniques to remove B-mode polarization arising from lensed E-modes.

Here is where we stand:  B-mode polarization from both dust and lensing has been detected and mapped out from $\ell \sim 40 - 2000$, with a current upper limit, $r \leq 0.1$; see Fig.~3.  The importance of the measurement is reflected in the very simple relationship between $r$ and the value of the inflationary potential and Hubble constant during inflation:  
\begin{eqnarray}
 V^{1/4} & = & 10^{16}\,{\rm GeV}\,\left( r\over 10^{-2} \right)^{1/4} \\
 H_{\rm inflation}^{-1} & = & 2 \times 10^{-38}\,{\rm sec}\,\left( r\over 10^{-2} \right)^{-1/2}
 \end{eqnarray}

\begin{figure}
\includegraphics[width=0.7\textwidth]{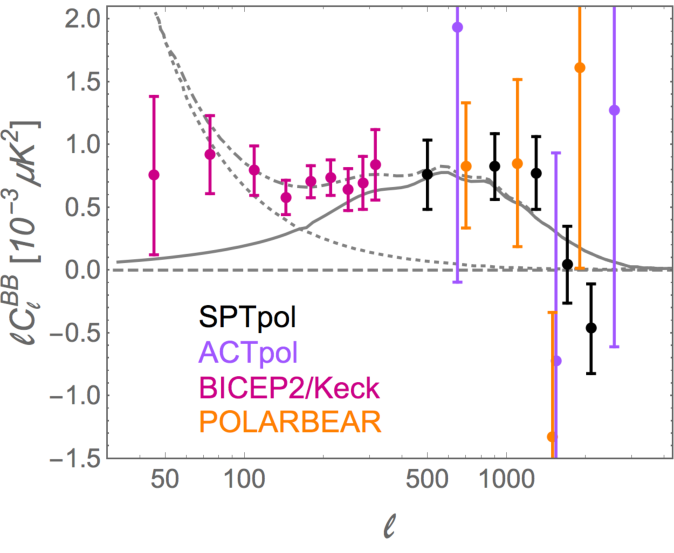}
\includegraphics[width=0.7\textwidth]{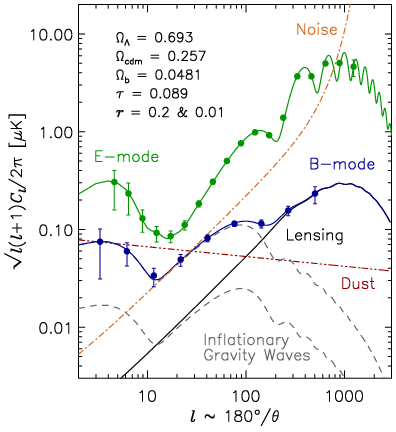}
\caption{Top:  B-mode power spectrum including theoretical expectations for $\rm\Lambda$CDM with $r=0$:  lensing (solid line), dust (dotted) and  total (broken) \cite{SPTBBmode}.  Bottom:  E-mode spectrum, B mode measurements and predictions for $r = 0.2$ and $= 0.01$ \cite{Araujo}}
\label{fig:3}       % Give a unique label
\end{figure}

\subsection{$H_0$:  disruption (and discovery?)}  There is currently a more than 3$\sigma$ discrepancy between direct measurements of the Hubble constant, using type Ia SNe calibrated by the astronomical distance scale, and measurements of $H_0$ inferred from CMB measurements, assuming $\rm\Lambda$CDM:  $73\pm 1.7$\,km/s/Mpc (direct) vs $67\pm 0.5$\,km/s/Mpc (CMB); see Fig.~4 \cite{Freedman}.  There could be an error in one or both measurements; or, they could both be correct and the problem is with the assumption of $\rm\Lambda$CDM.  If it is the latter, $\rm\Lambda$CDM is disrupted and possibly something new is discovered (e.g., neutrino mass or dark energy beyond $\rm\Lambda$).  At the moment, there is no compelling theoretical explanation of the two different values waiting to be verified, so a discovery would be an even bigger surprise.

\begin{figure}
\includegraphics[width=0.7\textwidth]{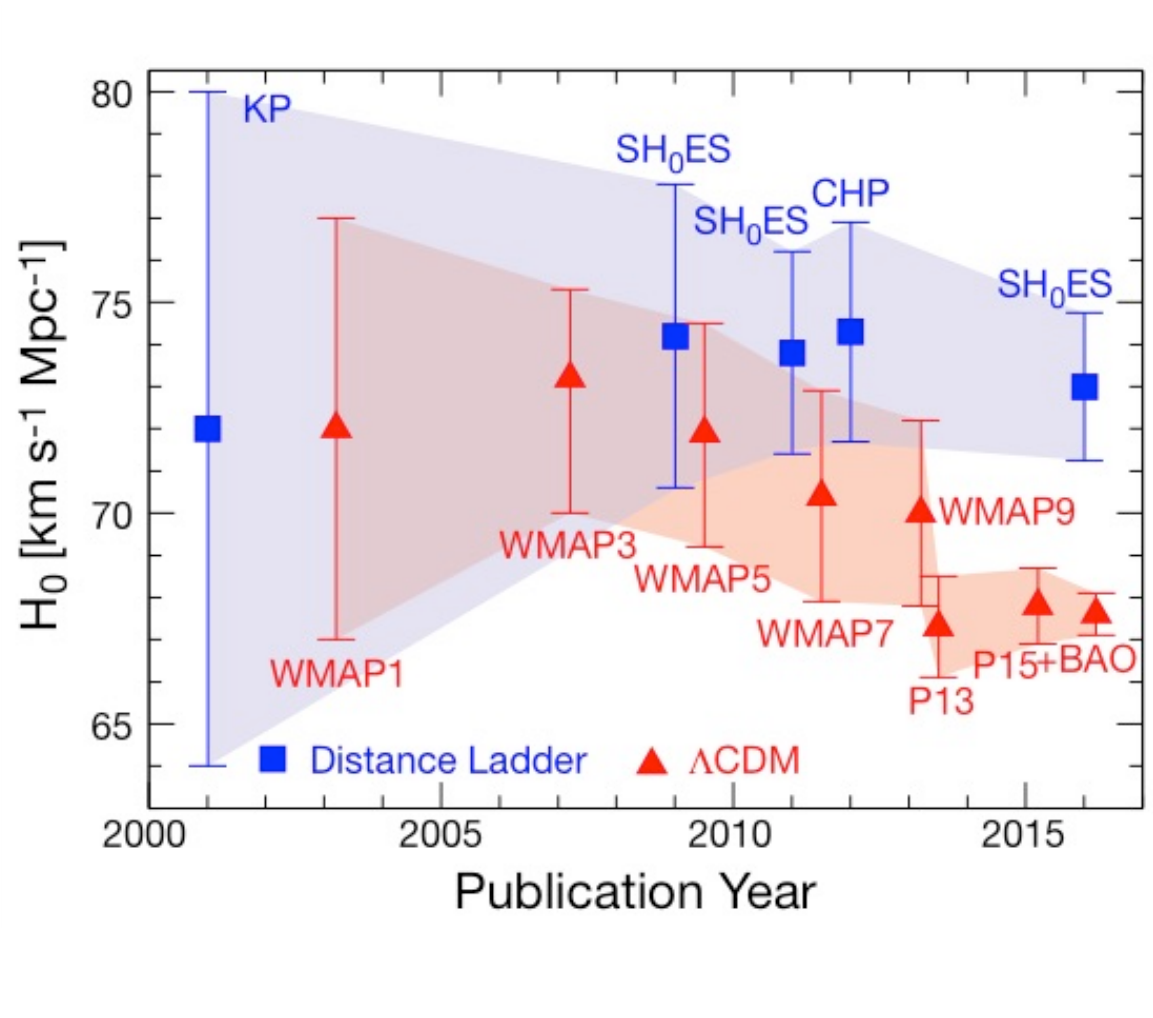}
\caption{$H_0$ tension:  Direct determinations in blue and CMB determinations in red differ by more than $3\sigma$ \cite{Freedman}.}
\label{fig:4}       % Give a unique label
\end{figure}

Allan Sandage must be happy:  Just when we thought the Hubble constant had nothing more interesting to tell us, this discrepancy reminds us that it is the most important number in cosmology!  One promising new approach to determining $H_0$ is gravitational-wave standard sirens:  the gravitational waveform of coalescing black holes (recently discovered!) or neutron stars determine the distance to the coalescing binary.  {\rm If} an electromagnetic counterpart to the event is detected and the redshift can be measured, a single event gives a direct determination of $H_0$ that is independent of the astronomical distance scale.\footnote{A few months after this meeting, the first detection of, and determination of the Hubble constant by, a standard siren was made by the LIGO/VIRGO Collaboration \cite{abbott}.}

\subsection{$\Lambda$ or GR:  disruption}  $\rm\Lambda$CDM assumes the correctness of General Relativity with a cosmological constant as the dark energy.  A measurement of $w \not= -1$ or a change in time ($w_a \not= 0$) would falsify $\rm\Lambda$ and disrupt the paradigm; measurements from the Dark Energy Survey, or the upcoming Euclid or WFIRST missions have the potential to do this.  A new technique, the growth of cosmic structure, has the additional benefit of being able to test the validity of the GR as the underlying theory of gravity.  In both cases, the absence of an compelling theoretical alternative makes it unlikely the disruption would also lead to a clear-cut discovery.

\subsection{Dark matter: discovery}  As mentioned in \S\ref{sec:2}, there are ambitious experimental programs searching for evidence of axions, neutralinos and WIMPs.  They are improving in sensitivity and already have sufficient sensitivity to cover interesting parts of the theoretical parameter space.  My favorite scenario is a triple detection of a WIMP or neutralino -- direct detection, indirect detection through annihilation products and accelerator production.  Not only would that confirm the particle dark matter hypothesis, but it would also allow for redundancy checks (e.g., consistency of the relic abundance with its measured annihilation cross section).

\subsection{CDM:  disruption}  CDM is the theory that has at least nine lives!  Since its introduction in 1983,  there have been many attempts to kill it, all unsuccessful.  (The only significant change in the paradigm came with the addition of $\rm\Lambda$, lowering the matter content to about 30\% of the critical density.)  CDM is a strong theory in the sense of Karl Popper:  it makes many predictions that can be used to falsify it.  Two current problems that are frequently mentioned are that the galactic rotation curves it predicts are too cuspy and that it predicts too many small satellites.  There is much controversy about how serious these problems are \cite{Silk}, and because baryonic `gastrophysics' can be important on these small scales it will be hard to rule out CDM this way.  Further, none of the alternatives, e.g., keV sterile neutrino dark matter or self-interacting dark matter, are compelling. 

Finally, because the successes of CDM are so numerous, especially on large scales where gastrophysics cannot confuse the situation, at the very least, CDM must have much of the truth, and the correct theory must be very similar to it.  Nonetheless, because CDM is a strong theory, we should keep an open mind to the possibly of its disruption.

\subsection{CMB anomalies:  disruption} Precision measurements of the anisotropy of the CMB have dramatically advanced cosmology and played a major role in establishing $\rm\Lambda$CDM.  However, they have also revealed some anomalies:  small quadrupole; correlations between the first few multipoles; and a striking `dent' around $\ell \sim 20 - 40$; see Fig.~1.  We have but one CMB sky to infer the underlying distribution and so sample variance -- here known as cosmic variance -- introduces an irreducible uncertainty.  Because of the absence of an {\it a priori} prediction, the first two anomalies reveal little because of significant cosmic variance.  The dent could be different and was discussed at this meeting by Bouchet \cite{Bouchet}.

\section{Some grand challenges beyond $\rm\Lambda$CDM}
\label{sec:4}
In the previous sections I have focused on the three pillars of $\rm\Lambda$CDM because of their central importance to the paradigm itself and their ripeness for testing.  Here, I mention some grander aspirations for cosmology.  Big questions, whose answers may not yet be within our reach.  And needless to say, my list is idiosyncratic and incomplete.

\subsection{Baryogenesis}  The revolution in cosmology initiated by bringing in ideas from particle physics began with baryogenesis in 1978.  At the time, baryogenesis, while attractive, was optional:  one could still imagine a net baryon number as an initial condition.  Once inflation is in the mix, the tremendous entropy produced by the re-heating process makes baryogenesis required because any initial baryon asymmetry would be reduced exponentially.  While we have compelling -- and testable -- models for the origin of dark matter, we can't say the same for the baryon asymmetry needed to ensure the existence of ordinary matter.  

The standard model has two of the necessary ingredients:  $B$ violation;  $C$ and $CP$ violation, and the expansion of the Universe can provide the needed departure from thermal equilibrium.  Electroweak baryogenesis once looked promising, but appears to fail on two accounts:  not enough $CP$ violation in the standard model and the dynamics of the phase transition do not provide the requisite departure from equilibrium.  Currently, much attention is focused on lepto/baryogenesis, where a lepton asymmetry is produced early on and then transformed into a baryon asymmetry by $B + L$ violating electroweak processes.  In either case, it appears that baryogenesis involves physics at scales that are inaccessible in the laboratory, making progress and testing difficult.  The grand challenge is a detailed model of baryogenesis that is testable.

\subsection{Who ordered that?!\#}  The history of cosmology is growing complexity in the  composition of the Universe:  baryons (circa 2500BCE), photons (1965), neutrinos (1967),  exotic dark matter (1981), cold dark matter (1983), massive neutrinos (1998), and dark energy (1998).   

In I.I. Rabi's words, who ordered all that?  Are there more relics?  What's next?   The cosmic mix raises interesting questions:  why is the ratio of CDM to baryons about 5?  and the ratio of baryons to neutrinos about 10?  (both ratios are of course constant in time) \cite{carrturner}.  Why did dark energy come to dominate the energy density so recently?  The grand challenge is explaining the mix and/or finding more relics.  

\subsection{Inflation:  up or out!}  Inflation has moved cosmology forward for more than three decades, but it is an incomplete theory and has teased us with some very big ideas, e.g., the multiverse.  The grand challenge  is simple:  a fundamental theory that makes sharp predictions that can be falsified, or an even better theory to replace it.

\subsection{Making sense of the multiverse}  The multiverse presents us with a big dilemma:  it may be the most important idea in science since Copernicus removed Earth from the center of the Universe, but it is inherently untestable,  and so isn't science.  I say that it is inherently untestable because the fundamental idea behind the multiverse is that the different pieces of the multiverse are causally distinct, which makes direct testing not just hard, but impossible in principle.  There are other important issues to be addressed to make sense of the multiverse, including that of defining a measure, but for me, the issue of whether or not it is within the realm of science is the go/no-go question.  The grand challenge is to make the multiverse science or to stop talking about it!

\section{Limits of scientific cosmology}
\label{sec:5}
Cosmology is an enterprise that requires hutzpah:  the Universe is enormous and often beyond the reach of our instruments and ideas.  Armed with no small amount of arrogance, bold ideas and powerful instruments, we have been remarkably successful in determining  the basic features of the Universe and reconstructing much of its history.  We hope to do even more.  However, there are limits to what we can learn.

\subsection{Destiny}  For years, we were confident that geometry and destiny were linked and that we could determine the fate of our Universe, cf. \cite{Sandage200inch}.  Cosmic acceleration and dark energy have humbled us.  We know the geometry -- flat to within 0.5\% -- but because of dark energy the question of destiny is even more unsettled.  First, there is a new possibility:  the big rip.  If $w < -1$, dark energy will lead to a singular ending in a finite amount of time.  Even if dark energy is explained by a scalar field that eventually relaxes to the minimum of its potential, returning the Universe to a new, dark matter-dominated era, we still cannot predict the destiny unless we solve the cosmological constant problem.  There could be a small -- really tiny -- cosmological constant of either sign that would change the outcome.  The destiny question is likely to remain open for a long time.

\subsection{Our past light cone}  G.F.R. Ellis has emphasized how the smallness of our past light cone severely limits our knowledge of the entire space-time \cite{Ellis}.  While invoking the Cosmological Principle may sound good, it is not a substitute for observational facts.  Moreover, if inflation is correct, the large-scale structure of the Universe could be very, very different from what we see in our observable Universe.  Because the Universe has undergone at least two periods of accelerated expansion, even the question of our past light cone is not a simple one.  The small fraction of space-time occupied by our past light cone fundamentally limits what we can learn about the Universe.

\subsection{The opaque screen of inflation}  A -- perhaps {\it the} -- fundamental theoretical motivation for something like inflation was to lessen the dependence of the present state of the Universe upon initial data.  Inflation has done its job well, maybe too well:  The entropy produced during reheating exponentially dilutes the abundance of any potential relic from the pre-inflation era; the accelerated expansion guarantees that the Universe will be appear to be flat, pushes any initial large-scale inhomogeneity outside our current horizon and impresses new density perturbations on small scales (i.e., those accessible today).  In short, if the Universe inflated, we have lost access to its initial state, and inflation has put up a very opaque screen that prevents us from seeing further back than the epoch of inflation.

\subsection{Multiverse}  Enough said earlier.  
%%If we live in a multiverse -- which has many attractive features -- we are unlikely to be able to verify such.

\subsection{Closing the circle}  Cosmology faces the challenge of being a `historical science,' concerned with reconstructing the timeline of the Universe (both backward and forward), and thus we cannot do experiments to directly test the timeline.  Of course, cosmology also allows us to use the `Heavenly Laboratory' to make discoveries beyond the reach of terrestrial laboratories -- e.g., the discoveries of $^4$He, dark matter and dark energy -- that can be later followed up on Earth.

Establishing the evolution of the Universe is not impossible, but can be very challenging.  The process typically involves a theoretical framework with testable predictions and observational evidence for those predictions, with the level of confidence depending upon both. The theoretical framework may involve theories that range from well-tested  to highly speculative.  Few would argue that the expansion of the Universe is not well-established and at the other end of the spectrum we have inflation, which is supported by some evidence, but  far from being well-established.  

I consider big-bang nucleosynthesis to be `well-established' and  it illustrates my `closing the circle' criterion.  The theoretical framework includes General Relativity (and its FLRW model for which there is much evidence),  nuclear physics, and statistical mechanics.  All three elements of the theoretical framework are well-tested in the laboratory; e.g., the crucial cross sections for BBN are all measured at the energies of relevance.  Further, BBN makes predictions that can be tested:  the primordial abundances of D, $^3$He, $^4$He and $^7$Li.  Within the standard model of particle physics, the predictions depend upon one cosmic parameter:  the baryon-to-photon ratio $\eta$, which is directly related to $\Omega_B$.  

With the exception of $^7$Li,\footnote{$^7$Li is a complicated and continuing story.  The theoretical predictions have the largest error bars and it is not clear that the primordial abundance has been determined.}  the predicted BBN light-element abundances agree with the measured primordial abundances for a single value of $\eta$, one which is consistent with the other determinations of the baryon density, most notably that from the CMB.  In spite of the fact that BBN involves events that occurred when the Universe was more than a billion times smaller and a billion times hotter, seconds after the beginning of the Universe, because the underlying physics is well established in the laboratory and its predictions have been verified, most cosmologists consider BBN a well-established part of the timeline of the Universe. It is the connection to physics well-established in the laboratory and verified predictions that closes the circle.

Inflation presents a bigger challenge.  On the one hand, two of the key predictions of the inflationary paradigm -- flat Universe with almost scale-invariant, Gaussian density perturbations -- have been confirmed, and great effort is being put forward to testing the third prediction, gravitational waves.  On the other hand, there is no standard model of inflation, some models have their roots in very speculative physics (e.g., superstring theory), and others are simply {\it ad hoc}. 

To illustrate the challenge inflation faces, let's suppose that the B-mode polarization signature of inflation-produced gravitational waves is detected.  This determines the Hubble constant during inflation, and allows the reconstruction of the inflationary potential.  So, would I include inflation in the well-established history of the Universe?  No:  Without a fundamental model of inflation rooted in laboratory experiments (e.g., the production of the inflaton particle at an accelerator), I believe the strongest statement I could make is that inflation holds much of the truth about the earliest moments of the Universe.

Finally, a few quick words about dark matter.  Until we discover the dark matter particle, dark matter and CDM  occupy a similar position:  a very attractive description of the formation of structure in the Universe that has much of the truth.  A triple detection of the dark matter particle would change that and close the circle, bringing CDM and particle dark matter into the well established history of the Universe.

%%\paragraph{Initial conditions}

\subsection{Limits of science}  As I like to remind my students (and the general public), science is the most powerful means we have found to explore the physical world.  Science has not only helped us understand the physical world and the laws that govern it, but it has also allowed us to use the physical laws for the betterment of humanity.  

However, there are limits to science.  It is all about how and not about why.  While we can aspire to discover the laws that govern the Universe and understand how they have shaped its evolution, science cannot explain why the laws are what they are, where they came from, or why we have something rather than nothing. Some questions are simply beyond the reach of science.  

\subsection{Final words}  No need to worry, there is plenty more that science can do in cosmology; $\rm\Lambda$CDM has only whet our appetite for more by allowing us to frame even more profound questions about the Universe.

%%Text with citations \cite{RefB} and \cite{RefJ}.
%%\subsection{Subsection title}
%%\label{sec:2}
%%as required. Don't forget to give each section
%%and subsection a unique label (see Sect.~\ref{sec:1}).
%%\paragraph{Paragraph headings} Use paragraph headings as needed.
%%\begin{equation}
%%a^2+b^2=c^2
%%\end{equation}


%\begin{figure*}
  %%\includegraphics[width=0.75\textwidth]{example.eps}
%\end{figure*}
%%\begin{table}
%%\caption{Please write your table caption here}
%%\label{tab:1}       % Give a unique label
%%\begin{tabular}{lll}
%%\hline\noalign{\smallskip}
%%first & second & third  \\
%%\noalign{\smallskip}\hline\noalign{\smallskip}
%%number & number & number \\
%%number & number & number \\
%%\noalign{\smallskip}\hline
%%\end{tabular}
%%\end{table}

\begin{acknowledgements}
This work was supported in part by the Kavli Institute for Cosmological Physics at the University of Chicago through grant NSF PHY-1125897 and an endowment from the Kavli Foundation and its founder Fred Kavli.
\end{acknowledgements}



\end{document}